\documentclass[11pt,a4paper]{article}
\usepackage{jcappub}

\usepackage{bm}
\usepackage{amsfonts}
\usepackage{subfigure}

\newcommand{\gsim}{\mbox{\;\raisebox{.3ex}
  {$>$}$\!\!\!\!\!$\raisebox{-.9ex}{$\sim$}}\;}

\newcommand{\be}{\begin{equation}}
\newcommand{\ee}{\end{equation}}
\newcommand{\bea}{\begin{eqnarray}}
\newcommand{\eea}{\end{eqnarray}}

\renewcommand{\vec}[1]{{\bm #1}}
\begin{document}

%%%%%%%%%%%%%%%%%%%%%%%%%%%%%%%%%%%%%%%%%%%%%%%%%%%%%%%%%%%%%%%%%%%%%%
% Frontpage %%%%%%%%%%%%%%%%%%%%%%%%%%%%%%%%%%%%%%%%%%%%%%%%%%%%%%%%%%
%%%%%%%%%%%%%%%%%%%%%%%%%%%%%%%%%%%%%%%%%%%%%%%%%%%%%%%%%%%%%%%%%%%%%%

%\subheader{\hfill Preprint ....}

\title{Updated constraints on non-standard neutrino interactions from Planck}

\author[a]{Maria Archidiacono}
\author[a]{Steen Hannestad}

\affiliation[a]{Department of Physics and Astronomy\\
 University of Aarhus, DK-8000 Aarhus C, Denmark}

\emailAdd{archi@phys.au.dk,sth@phys.au.dk}

\abstract{We provide updated bounds on non-standard neutrino interactions based on data from the Planck satellite as well as auxiliary cosmological measurements. Two types of models are studied - A Fermi-like 4-point interaction and an interaction mediated by a light pseudoscalar - and we show that these two models are representative of models in which neutrinos either {\it decouple} or {\it recouple} in the early Universe.
Current cosmological data constrain the effective 4-point coupling to be $G_X \leq \left(0.06 \, {\rm GeV}\right)^{-2}$, corresponding to $G_X \leq 2.5 \times 10^7 G_F$.
For non-standard pseudoscalar interactions we set a limit on the diagonal elements of the dimensionless coupling matrix, $g_{ij}$, of $g_{ii} \leq 1.2 \times 10^{-7}$. For the off-diagonal elements which induce neutrino decay the bound is significantly stronger, corresponding to $g_{ij} \leq 2.3 \times 10^{-11}(m/0.05 \, {\rm eV})^{-2}$, or a lifetime constraint of $\tau \geq 1.2 \times 10^{9} \, {\rm s} \, (m/0.05 \, {\rm eV})^{3} \,$. This is currently the strongest known bound on this particular type of neutrino decay.
We finally note that extremely strong neutrino self-interactions which completely suppress anisotropic stress over all of cosmic history are very highly disfavored by current data ($\Delta \chi^2 \sim 10^4$).} 

\maketitle

\section{Introduction}

It is well known that astrophysics and cosmology are excellent laboratories for probing neutrino physics. Precision measurements of the CMB anisotropy by the WMAP and Planck satellites as well as ground based experiments have, in combination with large scale structure surveys, allowed for quite stringent constraints on parameters such as the neutrino mass and energy density in the Universe (see e.g.\ \cite{Ade:2013zuv}).

Given the precision with which standard neutrino parameters such as the mass can be probed using measurements of cosmological structure formation it is of considerable interest to study how well neutrino interactions beyond the standard model can be tested (see e.g. \cite{AtrioBarandela:1996ur}). Two of the most often discussed versions of non-standard interactions are: (a) A new Fermi-like 4-point interaction mediated through a new vector boson, $X$, with an interaction strength given by $G_X = g^2/m_X^2$ with $g$ being the fundamental dimensionless coupling of the interaction. (b) A pseudoscalar interaction mediated through the Nambu-Goldstone boson, $\phi$, of a new broken $U(1)$ symmetry. The interaction strength in this case is simply given by the dimensionless coupling $g$ and we assume $\phi$ to be either exactly massless or at least light enough to always be relativistic for the system we look at here. 
While these two types of interactions by no means exhaust the possible model space they are actually surprisingly representative for the following reason: The 4-point model has a cross section rising steeply with energy and therefore $\Gamma/H$, where $\Gamma = n_\nu \langle \sigma |v| \rangle$ is the interaction rate, rises extremely fast with temperature. Interactions typically maintain equilibrium only as long as $\Gamma/H \gsim 1$ so that neutrinos are strongly coupled at early times and subsequently {\it decouple}. Qualitative this model is equivalent to the standard picture of neutrino interactions in the early universe apart from the possibility that $G_X$ is very different from $G_F$. Recently, this scenario has been investigated in Ref.~\cite{Cyr-Racine:2013jua}.
The pseudoscalar model (with massless pseudoscalars) has the intriguing feature that $\Gamma/H$ increases with temperature because $\Gamma \propto T$ and that neutrinos therefore {\it recouple}, i.e. they become strongly interacting at late times. Typically most models of non-standard neutrino interactions can be put into one of the two categories and therefore mapped to one of the two cases discussed here. In fact we will show that with the accuracy of current data the two models can be adequately represented by a simplified model. Case (a) can be mapped into a model in which neutrinos are infinitely strongly interacting until a redshift $z_i$ at which point they instantaneously become non-interacting. Likewise, case (b) can be mapped into a model where neutrinos are non-interacting until a redshift $z_i$ at which point they become infinitely strongly interacting. We note that this approximation to case (b) was also used in \cite{Basboll:2008fx}.

Finally we also note that the two cases studied here can be seen as extreme limiting cases of the same underlying interaction. Imagine an interaction mediated by a gauge boson with finite mass $M$. At temperatures much higher than $M$ the scattering cross section will have the same scaling with temperature as our pseudoscalar case, independent of the exact coupling structure. Conversely, at temperatures much lower than $M$ the interaction becomes point-like with a behavior equivalent to our Fermi-like interaction. We can therefore view the two cases as representative of a very wide range of possible neutrino interaction models in which the Fermi-like interaction represents the low energy effective field theory and the pseudo-scalar case the high energy limit of the UV-complete theory \cite{ArkaniHamed:2002sp}.

Case (b) is particularly interesting because it has dramatic effects at low energy, exactly the regime which can be tested using CMB \cite{Hannestad:2004qu,Cirelli:2006kt,Friedland:2007vv} and large scale structure data. As an example of how well cosmology constrains this type of model, the strongest known bound on invisible neutrino decays comes from considerations of cosmic structure formation \cite{Hannestad:2005ex,Serpico:2007pt,Basboll:2008fx}. Here we will update this bound using the latest cosmological data.
Conversely, case (a) typically dominates at high energy and it might be constrained by accelerator data. At low energy where $G_X^2 T^4$ is very small the 4-point interaction has typically decoupled unless $G_X$ is extremely large.

In this paper we study models of type (a) and (b) in order to establish cosmological bounds on the interaction strength. The paper is structured as follows: In section \ref{sec:boltzmannequations} we discuss the Boltzmann equation formalism used to describe interacting neutrinos in linear theory. In Section \ref{sec:4pointinteractions} we discuss non-standard interactions of the 4-point type, and in Section \ref{sec:pseudoscalarinteractions} pseudoscalar interactions are considered. We introduce our cosmological model in section \ref{sec:model}; in section \ref{sec:data} we list the data and explain the method used to constrain it and in section \ref{sec:results} we present our results. Finally, section \ref{sec:discussion} contains our conclusions.

\section{Boltzmann equations}
\label{sec:boltzmannequations}

The evolution of any given particle species in a dilute gas with no quantum entanglement can be followed using the single particle Boltzmann equation where the only coupling between species comes from possible scattering or annihilation processes. The notation used here follows that of Ma and Bertschinger \cite{ma} and uses the synchronous gauge (the equations could equally well be written in e.g.\ conformal Newtonian gauge).
As the time variable we use conformal time, defined as $d \tau = dt/a(t)$, where $a(t)$ is the scale factor. Also, as the momentum variable we shall use the
comoving momentum $q_j \equiv a p_j$. We further parametrize $q_j$ as
$q_j = q n_j$, where $q$ is the magnitude of the comoving momentum and $n_j$ is a unit 3-vector specifying direction.

The Boltzmann equation can generically be written as
\begin{equation}
L[f] = \frac{Df}{D\tau} = C[f],
\end{equation}
where $L[f]$ is the Liouville operator.
The collision operator on the right-hand side describes
any possible collisional interactions.

One can then write the distribution function as
\begin{equation}
f(x^i,q,n_j,\tau) = f_0(q) [1+\Psi(x^i,q,n_j,\tau)],
\end{equation}
where $f_0(q)$ is the unperturbed distribution function.

In synchronous gauge the Boltzmann equation can be written as an evolution equation for $\Psi$ in $k$-space \cite{ma}
\begin{equation}
\frac{1}{f_0} L[f] = \frac{\partial \Psi}{\partial \tau} + i \frac{q}{\epsilon}
\mu \Psi + \frac{d \ln f_0}{d \ln q} \left[\dot{\eta}-\frac{\dot{h}+6\dot{\eta}}
{2} \mu^2 \right] = \frac{1}{f_0} C[f],
\label{eq:boltzX}
\end{equation}
where $\mu \equiv n^j \hat{k}_j$ and $\epsilon = (q^2+a^2 m^2)^{1/2}$.
$h$ and $\eta$ are the metric perturbations, defined from the perturbed space-time
metric in synchronous gauge \cite{ma}
\begin{equation}
ds^2 = a^2(\tau) [-d\tau^2 + (\delta_{ij} + h_{ij})dx^i dx^j],
\end{equation}
\begin{equation}
h_{ij} = \int d^3 k e^{i \vec{k}\cdot\vec{x}}\left(\hat{k}_i \hat{k}_j h(\vec{k},\tau)
+(\hat{k}_i \hat{k}_j - \frac{1}{3} \delta_{ij}) 6 \eta (\vec{k},\tau) \right).
\end{equation}

\subsection{Collisionless Boltzmann equation}

Let us first study the Boltzmann equation in the collisionless limit, i.e.\ $\frac{1}{f_0} C[f] = 0$. 
The perturbation is then expanded as
\begin{equation}
\Psi = \sum_{l=0}^{\infty}(-i)^l(2l+1)\Psi_l P_l(\mu).
\end{equation}
Following \cite{ma} we can now write the Boltzmann equation as a moment hierarchy for the $\Psi_l$
by performing the angular integration of $L[f]$
\begin{eqnarray}
\dot\Psi_0 & = & -k \frac{q}{\epsilon} \Psi_1 + \frac{1}{6} \dot{h} \frac{d \ln f_0}
{d \ln q} \label{eq:psi0}\\
\dot\Psi_1 & = & k \frac{q}{3 \epsilon}(\Psi_0 - 2 \Psi_2) \label{eq:psi1}\\
\dot\Psi_2 & = & k \frac{q}{5 \epsilon}(2 \Psi_1 - 3 \Psi_3) - \left(\frac{1}{15}
\dot{h}+\frac{2}{5}\dot\eta\right)\frac{d \ln f_0}{d \ln q} \\
\dot\Psi_l & = & k \frac{q}{(2l+1)\epsilon}(l \Psi_{l-1} - (l+1)\Psi_{l+1})
\,\,\, , \,\,\, l \geq 3
\end{eqnarray}
It should be noted here that the first two hierarchy equations are directly
related to energy and momentum conservation respectively (see e.g. \cite{Hannestad:2000gt}).

\subsection{Collisional Boltzmann equation}

Starting from the general Boltzmann equation in synchronous gauge, Eq.~(\ref{eq:boltzX}) we can introduce interactions 
by lifting the restriction that $\frac{1}{f_0} C[f] = 0$. 
In general the right hand side will be complicated, reflecting the particular structure of the interaction. However, for our purpose here
it suffices to use an approximate relaxation time treatment, as has typically been done in most Boltzmann equation treatments of non-standard
neutrino interactions in the cosmological context.

As long as interactions are number conserving, i.e.\ elastic scattering processes the collision terms in the $l=0$ and $l=1$ equations are both
explicitly zero because of energy and momentum conservation (see e.g.\ \cite{Hannestad:2000gt}).
All the higher order terms can be estimated from the relaxation time approximation \cite{Hannestad:2000gt},
in which
\begin{equation}
\frac{1}{f_0}C[f] = -\frac{\Psi}{\tau}.
\label{eq:reltime}
\end{equation}
Here, $\tau$ is the mean time between collisions
\begin{equation}
\tau^{-1} = \Gamma = n \langle \sigma |v| \rangle,
\end{equation}
where $\langle \sigma |v| \rangle$ is the thermally averaged cross section.
We note that the right hand side of the Boltzmann equation in principle should be calculated using the full scattering kernel of the interaction considered. This would lead to a momentum dependent relaxation time, $\tau_l(q)$ for each moment in the hierarchy. However, since observables depend only on integrated quantities such as energy and momentum density we at most make a minor error by using the thermally averaged relaxation time, $\tau_l(q) \sim \tau_l$, for each multipole. For all $l$ higher than 1 the relaxation times should differ by at most numerical factors of $\mathcal{O}(1)$ and we therefore use only one effective relaxation time for all $l$, $\tau_l \sim \tau$.

\section{4-point interactions}
\label{sec:4pointinteractions}

This is the most often discussed version of non-standard interactions (see e.g.\ \cite{Ohlsson:2012kf} for a recent review). In principle it is simply parametrized in terms of $G_X$ for the neutrino-neutrino interaction, as it has been recently studied in Ref.~\cite{Cyr-Racine:2013jua} where the massless neutrino case is considered, while here we include massive neutrinos. However, since the underlying model has both a vector boson mass $m_X$ and a dimensionless coupling $g$ there are cases where it is not enough to worry about $G_X = g^2/m_X^2$. For example, if $G_X$ is very large, $m_X$ can become so small that there is a thermal population of these particles which needs to be taken into account. For the models studied here this will never be the case and we can simply treat the new interaction as an additional elastic scattering term for neutrinos.

\begin{figure}[h]
\centering
\includegraphics[scale=0.45]{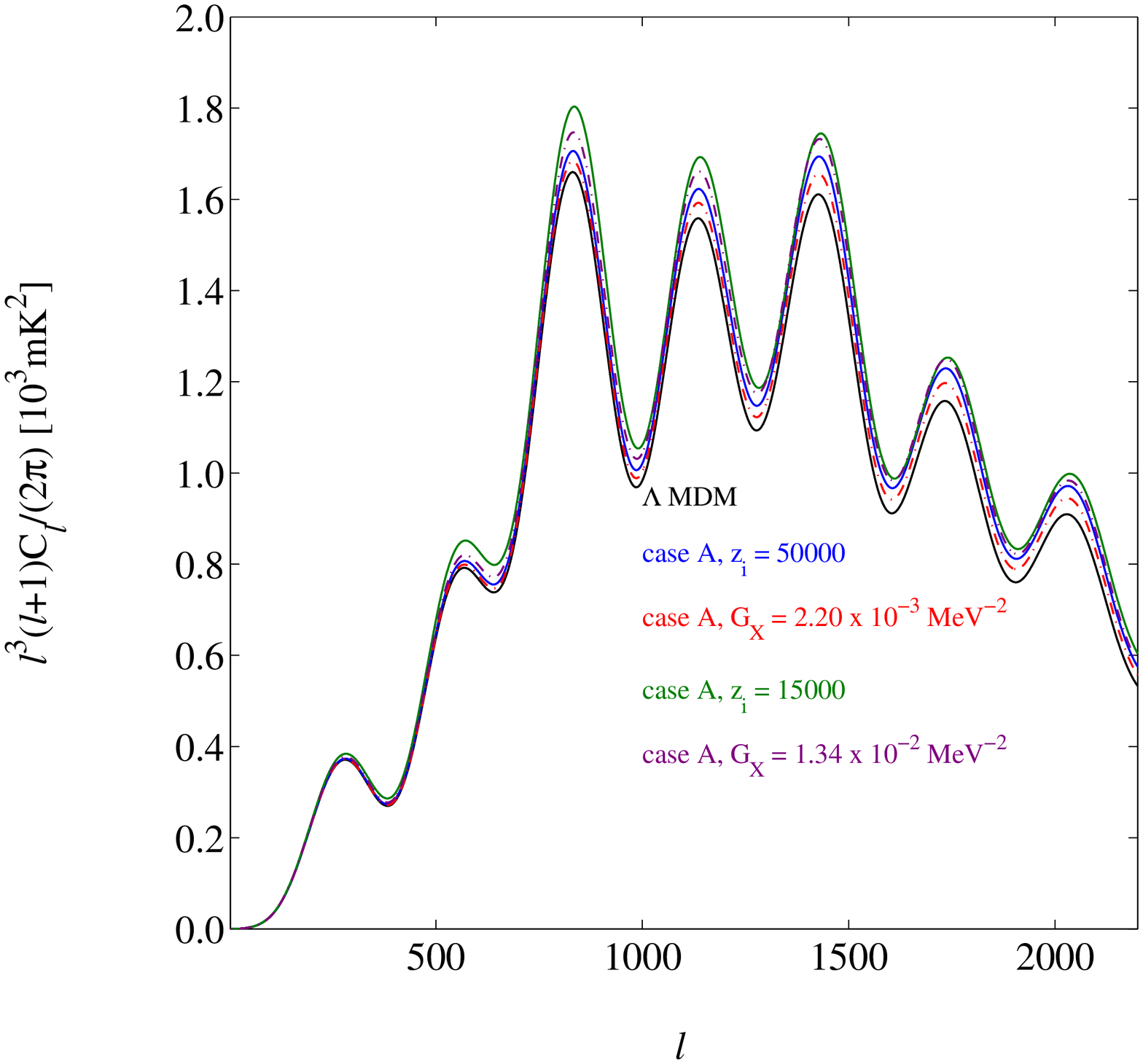} \\
\includegraphics[scale=0.45]{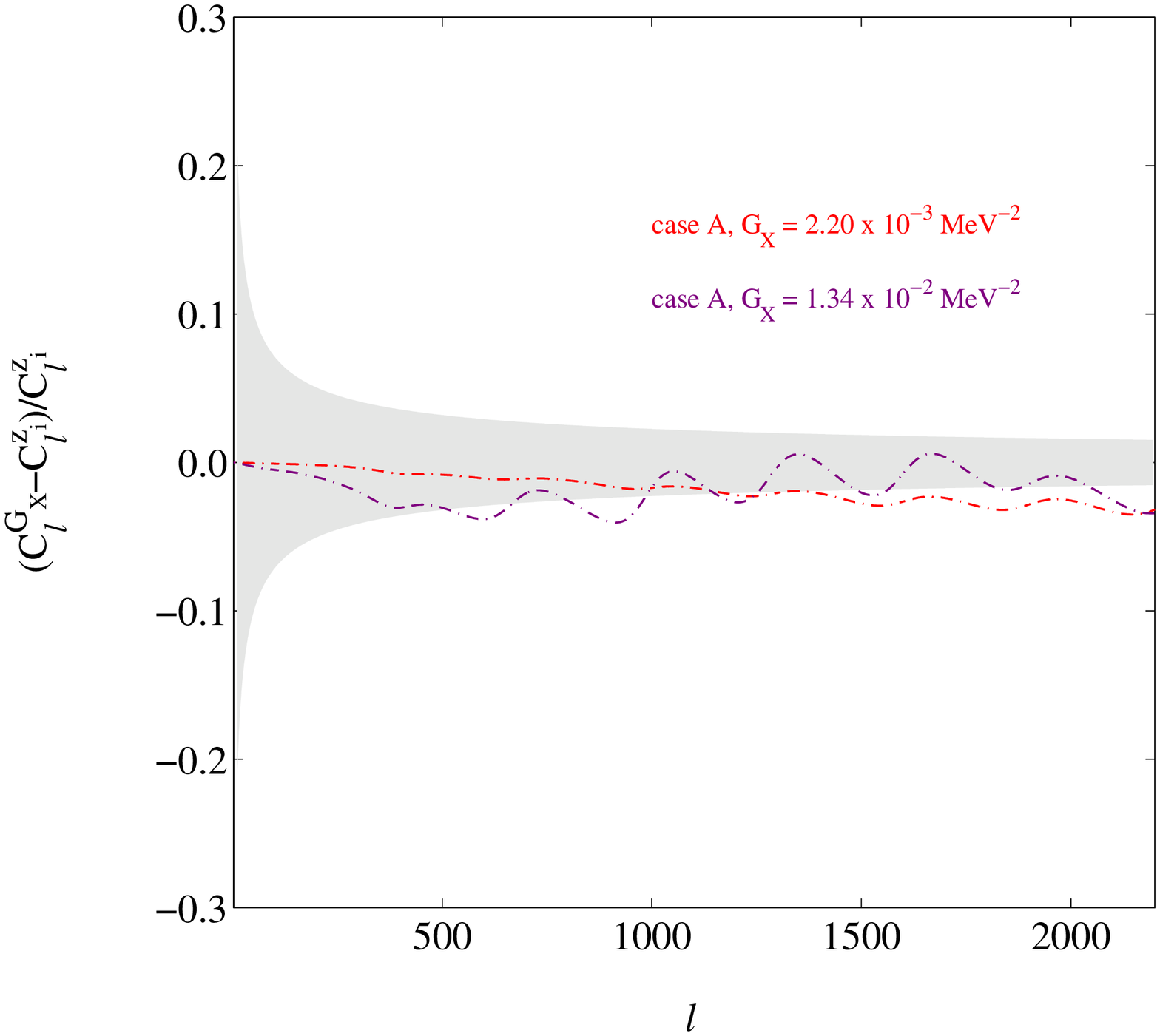} \\
\caption{{\bf case A} ({\it Top panel}) CMB temperature angular power spectra. The black line shows the Lambda Mixed (Cold + Hot) Dark Matter ($\Lambda$MDM) model with $\omega_{\rm cdm}=0.099$ and $\omega_\nu = 0.013$ (corresponding to $\Sigma m_\nu=1.2$~eV). The blue and green lines depict the theoretical spectrum obtained if massive neutrinos decouple at redshift $z_i=50000$ and $z_i=15000$, respectively. The red and purple lines represent the corresponding power spectrum obtained by plugging into the Boltzmann equations collisional term related to the $G_X$ value corresponding to the decoupling redshift found from Eq.~(\ref{eq:gtozcaseA}): $G_X=2.20\times10^{-3}\, {\rm MeV}^{-2}$ for a decoupling redshift $z_i=50000$ and $G_X=1.34\times10^{-2}\, {\rm MeV}^{-2}$ for a decoupling redshift $z_i=15000$. ({\it Bottom panel}) Percentage error introduced by the approximations of switching off the hierarchy at $z<z_i$ instead of plugging into the equations the correct expression of $G_X$. The grey band defines the cosmic variance.}
\label{fig:clstaufermi}
\end{figure}

For the 4-point interaction we have 
\begin{equation}
\langle \sigma |v| \rangle \simeq G_X^2 \begin{cases}T^2 & T \geq m_\nu \\ m_\nu ^{3/2} T^{1/2} & T \leq m_\nu \end{cases}.
\end{equation}
This means that since $\tau^{-1} = n \langle \sigma |v| \rangle$ we have $\tau^{-1} \propto T^5$ in the relativistic limit and we can to a good approximation have neutrinos be strongly interacting up to $z_i$ and free-streaming afterwards.
In Fig.~\ref{fig:clstaufermi} (top panel) we show the error introduced in the CMB power spectrum by using this approximation (blue and green solid line) rather than plugging into the Boltzmann equation the collisional term related to the expression of $G_X$ (red and purple dash-dotted line); the power spectrum of the $\Lambda$MDM model is also shown as reference. The percentage differences between the approximations (at redshift $z_i=50000$ and $z_i=15000$) and the spectra obtained with the corresponding couplings
($G_X=2.20\times10^{-3}\, {\rm MeV}^{-2}$ and $G_X=1.34\times10^{-2}\, {\rm MeV}^{-2}$, respectively) plotted in the bottom panel confirm that the approximation is reasonable if the decoupling redshift turns out to be greater than $\sim10^5$.
Assuming that we are in the radiation dominated epoch and that we can use the cross section in the relativistic limit, $z_i$ is related to $G_X$ approximately via
\begin{equation}
\frac{G_X}{(10 \, {\rm MeV})^{-2}} \sim \left(\frac{10^4}{z_i}\right)^{3/2}.
\label{eq:gtozcaseA}
\end{equation}

\subsection{Constraints from other sources}

Ref.~\cite{Bilenky:1999dn} provides a list of bounds on $G_X$ for 4-point interactions involving only neutrinos, as opposed to the more often studied non-standard interactions involving neutrinos and charged lepton or quark 4-point interactions (see e.g.\ \cite{Davidson:2003ha,Biggio:2009kv}). The tightest constraints are from 1-loop contributions to the invisible $Z$ decay width and leads to constraints of order $G_X \leq (1-10) \times G_F$, i.e.\ they apparently exclude the model studied here by many orders of magnitude. Somewhat less stringent bounds can be derived from $K$ decay. We also note that if the new boson couples to charged leptons the bounds are significantly stronger and in general rule out $m_X$ in the range where it could be relevant for CMB and structure formation (see e.g.\ \cite{Laha:2013xua}).
However, all these bounds rely on the 4-point description being valid at the energy scale of the experiment which is typically in the multi GeV range. If the mass of the new boson is substantially lower than this the bounds dilute substantially because additional factors of $m_X^2/m_Z^2 \ll 1$ will appear (for example the $Z$ decay correction is suppressed by approximately a factor $m_X^4/m_Z^4$). A very robust bound which is applicable at low energy is the one derived from neutrino observations of SN1987a \cite{Kolb:1987qy}. Here, it was found that $g/m_s < 12/{\rm MeV}$, a very loose bound which leaves room for interactions of the kind studied here. In summary it is not clear that any existing laboratory bounds actually exclude the 4-point interaction studied here, but in any case we can  take this scenario as a worked example of general models in which neutrinos decouple late.

\section{Pseudoscalar interactions}
\label{sec:pseudoscalarinteractions}

Another possible non-standard interaction is that neutrinos couple to a light pseudoscalar degree of freedom, $\phi$, via a Lagrangian of the form
\begin{equation}\label{eq:lagrangian}
{\cal L} = -i\,\phi\sum_{jk} g_{jk} \bar{\nu}_j \gamma_5 \nu_k\,,
\end{equation}
where the indices refer to neutrino states in the mass basis. Most likely the pseudoscalar arises from a broken $U(1)$ symmetry, as is the case in
e.g.\ majoron models \cite{Gelmini:1980re, Raffelt:1987ah, Barger:1981vd}. In that case a derivative coupling is more appropriate in certain cases. However, as was discussed in \cite{Hannestad:2005ex} using a derivative coupling
actually leads to more restrictive limits for the scenario studied here and we will use the pseudoscalar bounds as a conservative estimate.
Apart from numerical factors
the scattering rate in a thermal environment of relativistic neutrinos
is
\begin{equation}
\Gamma_{1+2\leftrightarrow3+4}\approx g^4 T\,,
\label{eq:binary}
\end{equation}
where $g$ can be approximated by largest entry of the Yukawa coupling matrix.

First of all, the process $\nu+\nu\to\nu+\nu$ leads to neutrino self-interactions which reduce anisotropic stress. However, the light pseudoscalar
can also be produced in processes such as $\nu+\bar\nu\leftrightarrow\phi+\phi$ and leads to energy and momentum transfer to an additional fluid. As long
as both neutrinos and pseudoscalars are relativistic they can be treated as one fluid with self-interactions. However, if neutrinos are massive they will
start transferring energy and entropy to $\phi$ as soon as $T < m_\nu$ and this effect must in principle be accounted for by tracking the evolution of the $\nu$
and $\phi$ fluids separately.
In \cite{Hannestad:2004qu} the effect was discussed in detail, and whether treating neutrinos and pseudoscalars as one fluid is a good approximation depends crucially on the mass of the neutrino.
As soon as neutrinos become non-relativistic they will pair-annihilate into $\phi$ particles, leading to the scenario known as the ``neutrinoless universe'' 
\cite{Beacom:2004yd}. 
In practise the bound on $g$ is so strong that neutrinos up to the eV mass scale are excluded from being strongly interacting prior to around recombination. Since the cosmological data we use are 
almost exclusive affected by neutrino evolution prior to recombination, treating neutrinos and pseudo-scalars as a single fluid is for our purposes a 
reasonable approximation. For the remainder of this work we shall be making this assumption.

However, the mass bounds derived in the next section turn out to restrict neutrinos to be so light that they are relativistic around recombination, justifying
our treatment of neutrinos and pseudoscalars as one fluid.

\begin{figure}[h]
\centering
\includegraphics[scale=0.45]{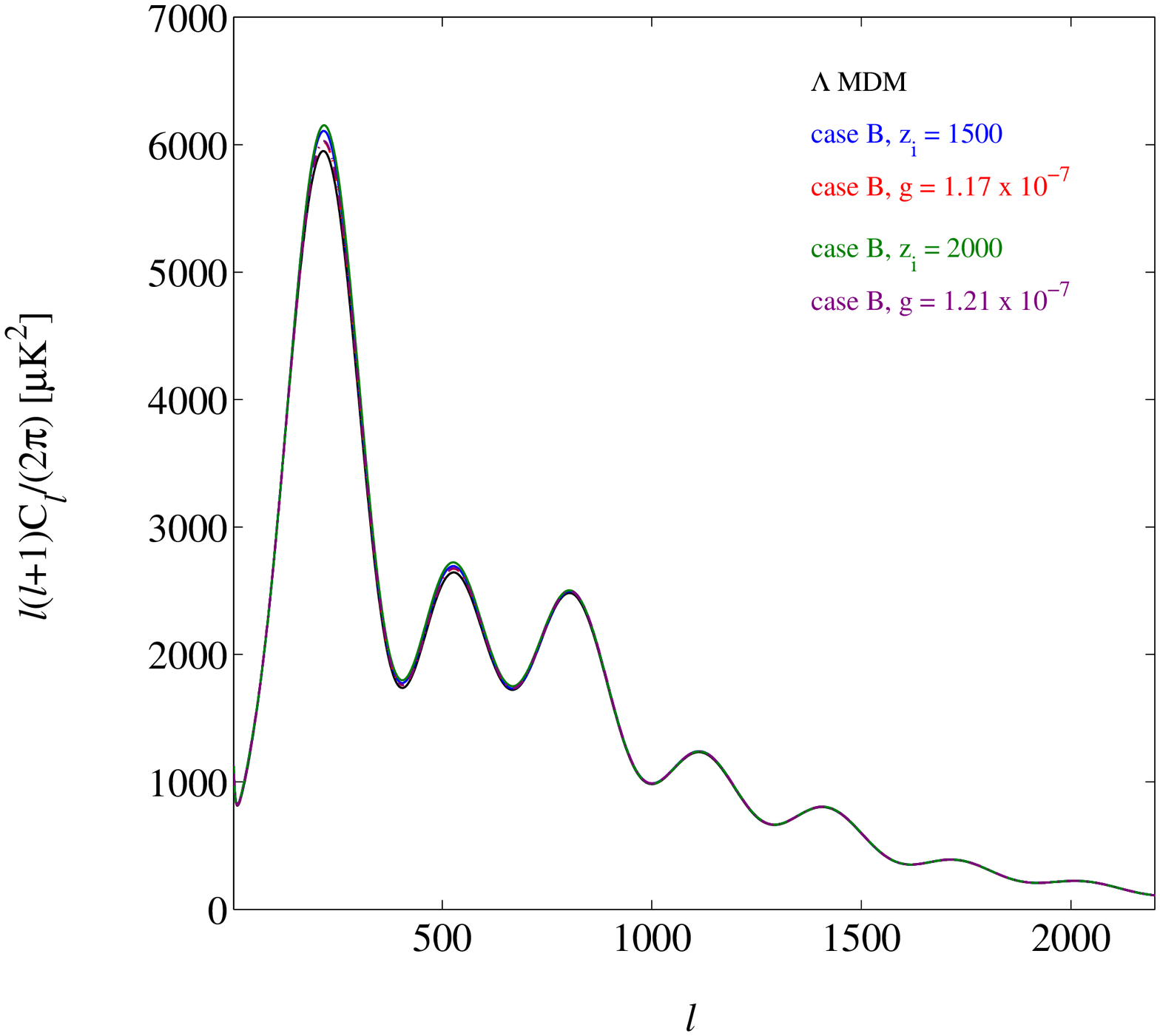} \\
\includegraphics[scale=0.45]{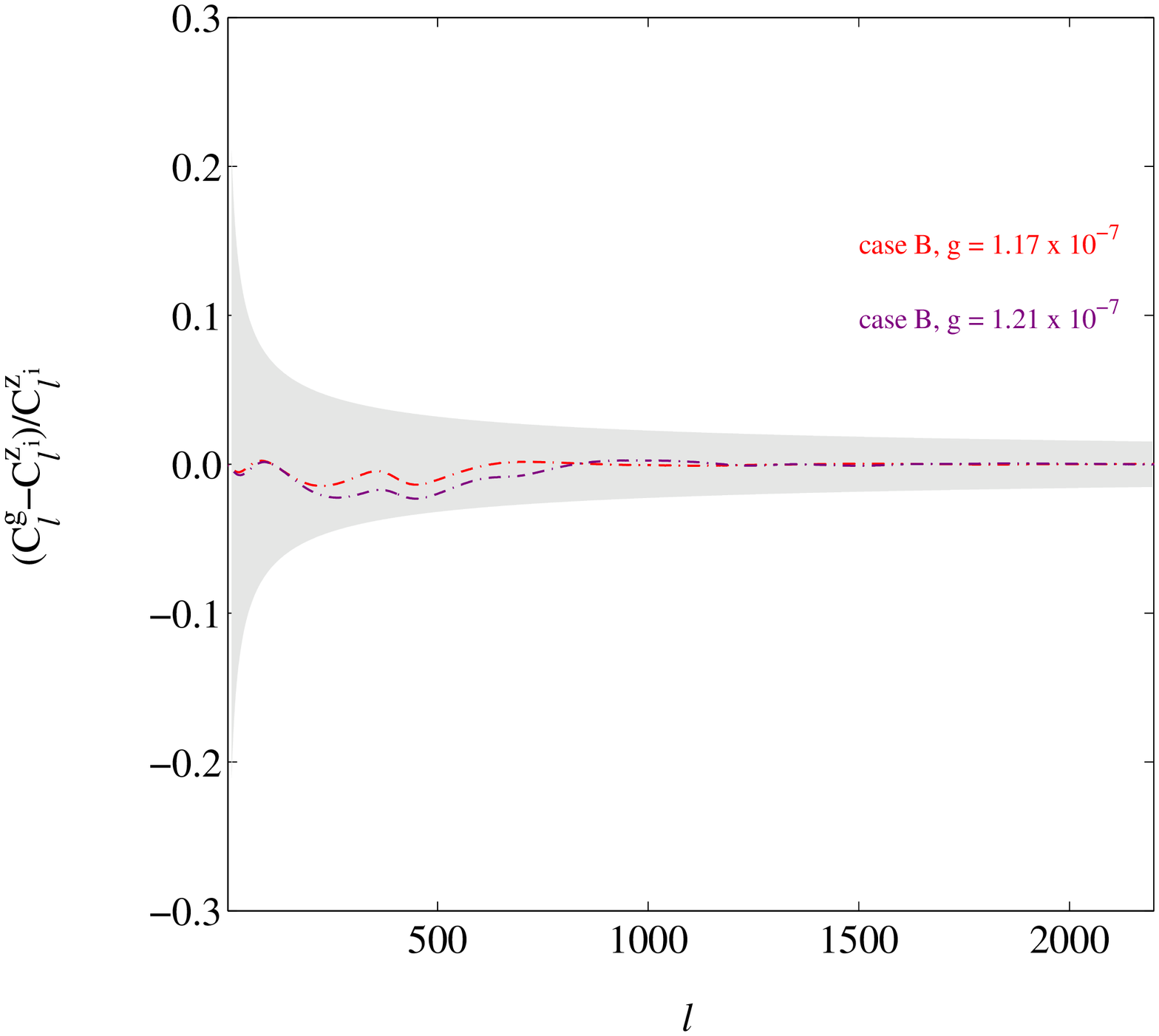} \\
\caption{{\bf case B} ({\it Top panel}) CMB temperature angular power spectra. The black line shows the $\Lambda$MDM model with $\omega_{\rm cdm}=0.099$ and $\omega_\nu = 0.013$ (corresponding to $\Sigma m_\nu=1.2$~eV). The blue and green lines depict the theoretical spectrum obtained if massive neutrinos recouple through the interactions with a pseudoscalar at redshift $z_i=1500$ and $z_i=2000$, respectively. The red and purple lines represent the corresponding power spectrum obtained by plugging into the Boltzmann equations the collisional term related to the $g$ value corresponding to the recoupling redshift found from Eq.~(\ref{eq:gtozcaseB}): $g<1.17\times10^{-7}$ when $z_i=1500$ and $g<1.21\times10^{-7}$ when $z_i=2000$. ({\it Bottom panel}) Percentage error introduced by the approximations of switching off the hierarchy at $z>z_i$ instead of plugging into the equations the correct expression of $g$. The grey band defines the cosmic variance.}
\label{fig:clstaupseudo}
\end{figure}

Using the same argument as for the case of the 4-point interaction we can approximate the pseudoscalar model with one where neutrinos are non-interacting
until a redshift $z_i$ after which they become strongly interacting. 
We test the correctness of this approximation in Fig.~\ref{fig:clstaupseudo}. In the top panel, together with the $\Lambda$MDM power spectrum, we show the power spectra obtained with the approximation (blue solid line $z_i=1500$ and green solid line $z_i=2000$) and with the corresponding upper limit of the coupling constant (red dash-dot line $g<1.17\times10^{-7}$ and purple dash-dot line $g<1.21\times10^{-7}$). The percentage error (bottom panel) introduced by the approximation remains within the cosmic variance limit. Nevertheless we notice that a variation of the recoupling redshift affects the spectrum the way more than the corresponding variation on the coupling constant.
Given the interaction rate in Eq.~(\ref{eq:binary}) we can approximate the relation between
$g$ and $z_i$ with
\begin{equation}
g \sim  1.1 \times 10^{-7}  \begin{cases} \left(\frac{z_i}{3000}\right)^{1/4} & {\rm RD} \\
\left(\frac{z_i}{1088}\right)^{1/8} & {\rm MD}
\end{cases}
\label{eq:gtozcaseB}
\end{equation}

Because of the $g^4$ dependence of the $2 \leftrightarrow 2$ interaction rate, 
corrections
to the simple expression in Eq.~(\ref{eq:binary}) which was derived by dimensional analysis are very small and can for all practical purposes be ignored. We can thus
treat the limits obtained in the next section as fairly robust.

The binary processes discussed above constrain any element of $g_{jk}$, including the diagonal ones. However, the off-diagonal terms induce 
decay processes like $\nu_j \to \nu_k \phi$, a process which is only ${\cal O}(g^2)$ and therefore at least in principle significantly more sensitive to 
$g$ than the ${\cal O}(g^4)$ binary processes.

The sum of the decay rates for $\nu\to\nu'+\phi$ and
$\nu\to\bar\nu'+\phi$ in the rest frame of the parent neutrino
with mass $m\gg m'$ is~\cite{Kim:1990km, Beacom:2002cb}
\begin{equation}\label{eq:decayrate}
\Gamma_{\rm decay}=\frac{g^2}{16\pi}\, m\,.
\end{equation}
In the frame of the thermal medium, a typical neutrino energy is
$E \sim 3T$ so that the rate is reduced by the corresponding Lorentz factor $m/3T$.

The phase space of decay and inverse decay processes is kinematically constrained
for relativistic particles and couples only nearly collinear modes of
the interacting particles.

Therefore, even if the decay is isotropic in the rest frame of the parent particle, the decay
products will have directions within an approximate angle, $\theta$,
corresponding to the Lorentz factor, $m/E$, of the parent particle. 
This introduces a difference between the decay rate and the rate with which direction of momentum
can be transferred in the system. The rate with which direction of momentum can be changed is
the one relevant for e.g.\ acoustic waves propagating in the system and therefore the
relevant rate constrained by observations of the CMB and baryon acoustic oscillations.
As was discussed in detail in \cite{Hannestad:2005ex} the difference between the two rates is approximately 
a factor $(m/E)^2$ and the relevant rate is therefore 

\begin{equation}
\Gamma_{\rm T}\approx\frac{g^2}{16\pi}\, m\,
\left(\frac{m}{E}\right)^3
\end{equation}

%If the coupling $g$ is between two nearly degenerate mass eigenstates,
%the decay rate acquires an additional factor of approximately $(\delta
%m^2)^3/m^6$~\cite{Beacom:2002cb} so that the limit on $g$ is relaxed
%by a factor $m^3/(\delta m^2)^{3/2}$.

When looking at bounds on the off-diagonal elements we shall again be using the same
technique as for the binary processes, i.e.\ assume that neutrinos are non-interacting until $z_i$ after which
they become infinitely strongly interacting.

\begin{figure}[h]
\centering
\includegraphics[scale=0.45]{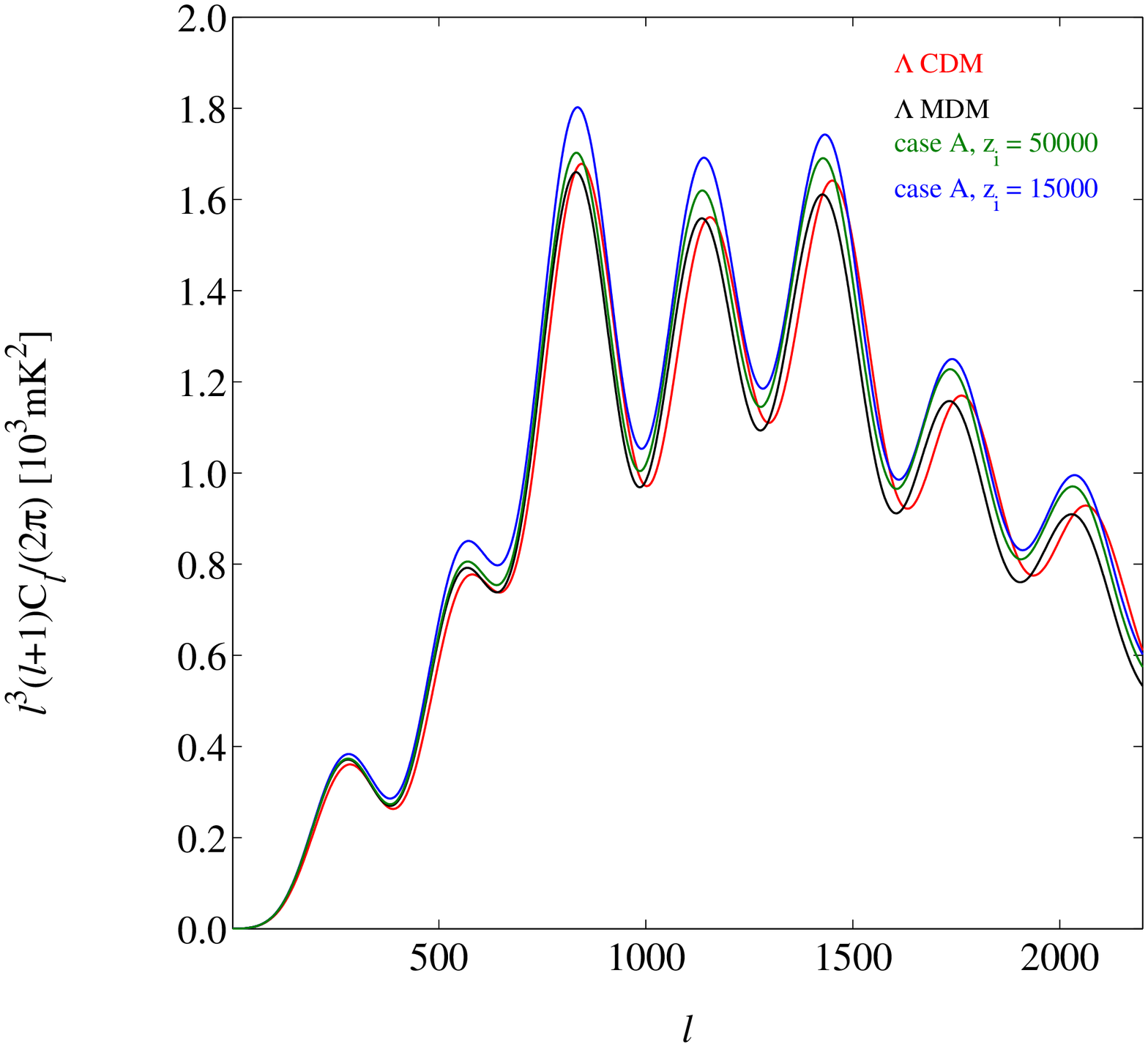} \\
\includegraphics[scale=0.45]{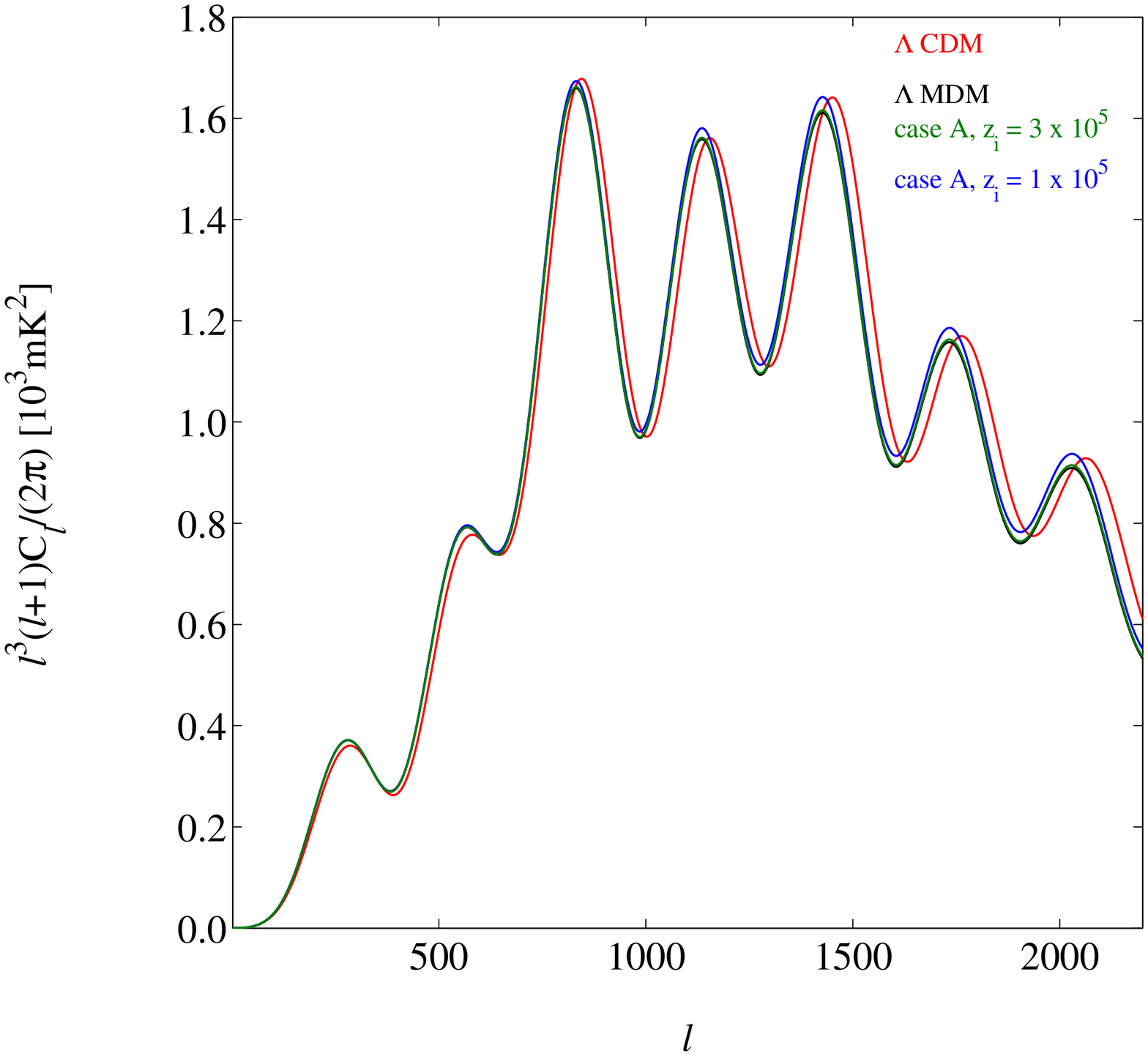} \\
\caption{{\bf case A} CMB temperature angular power spectrum for three different cosmological models: the red line shows the $\Lambda$CDM model with $\omega_{\rm cdm}=0.112$, while the black line is the $\Lambda$MDM model with $\omega_{\rm cdm}=0.099$ and $\omega_\nu = 0.013$ (corresponding to $\Sigma m_\nu=1.2$~eV). The blue and green lines depict the spectra obtained if Fermi like 4-point interactions prevent massive neutrinos from free-streaming until redshift $z_i=15000$ (blue line, upper panel), $z_i=50000$ (green line, upper panel), $z_i=1\times10^5$ (blue line, bottom panel), $z_i=3\times10^5$ (green line, bottom panel).}
\label{fig:clscaseA}
\end{figure}

\begin{figure}[h]
\centering
\includegraphics[scale=0.45]{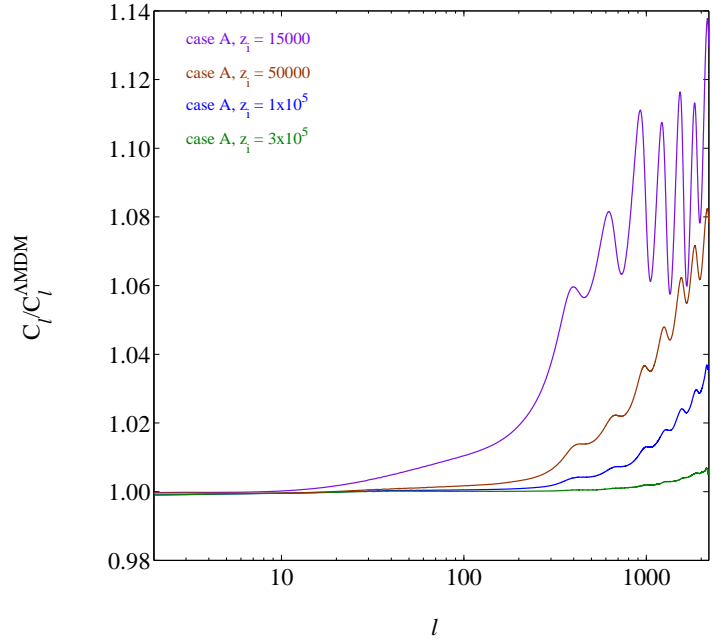} \\
\caption{{\bf case A} Ratio between the CMB temperature power spectra accounting for the Fermi like 4-point interactions at different redshift (as described in Fig.~\ref{fig:clscaseA}) and the $\Lambda$MDM spectrum.}
\label{fig:clsratiocaseA}
\end{figure}

\begin{figure}[h]
\centering
\includegraphics[scale=0.45]{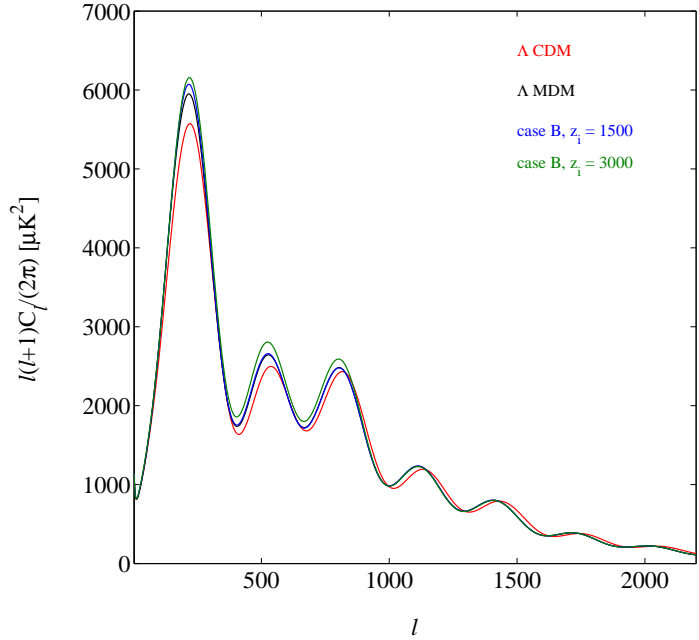}
\caption{{\bf case B} CMB temperature angular power spectrum for three different cosmological models: the red line shows the $\Lambda$CDM model with $\omega_{\rm cdm}=0.112$, while the black line is the $\Lambda$MDM model with $\omega_{\rm cdm}=0.099$ and $\omega_\nu = 0.013$ (corresponding to $\Sigma m_\nu=1.2$~eV). The blue and green lines depict the theoretical spectra obtained if massive neutrinos recouple through the interactions with a light pseudoscalar degree of freedom at redshift $z_i=1500$ or $z_i=3000$, respectively.}
\label{fig:clscaseB}
\end{figure}

\begin{figure}[h]
\centering
\includegraphics[scale=0.45]{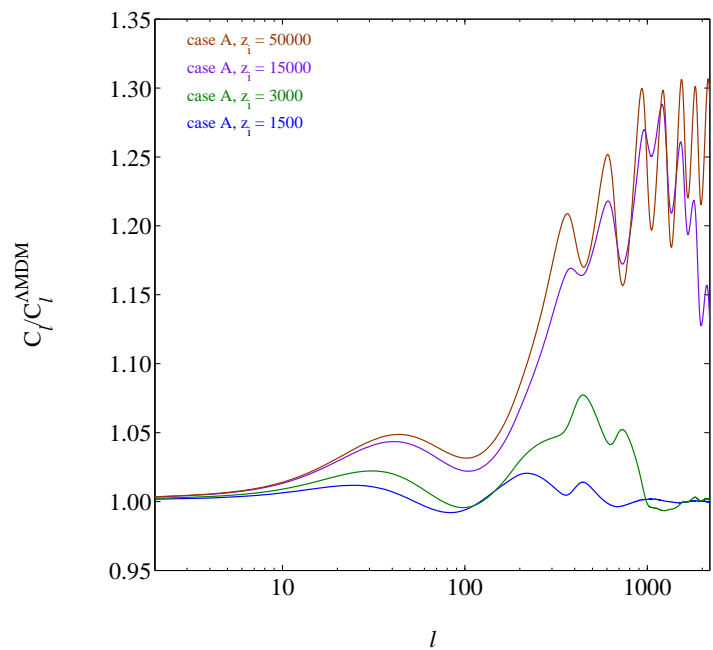} \\
\caption{{\bf case B} Ratio between the CMB temperature power spectra accounting for the pseudoscalar interactions at different redshift (as described in Fig.~\ref{fig:clscaseB}) and the $\Lambda$MDM spectrum.}
\label{fig:clsratiocaseB}
\end{figure}

\begin{figure}[h]
\centering
\begin{tabular}{cc}
\includegraphics[scale=0.3]{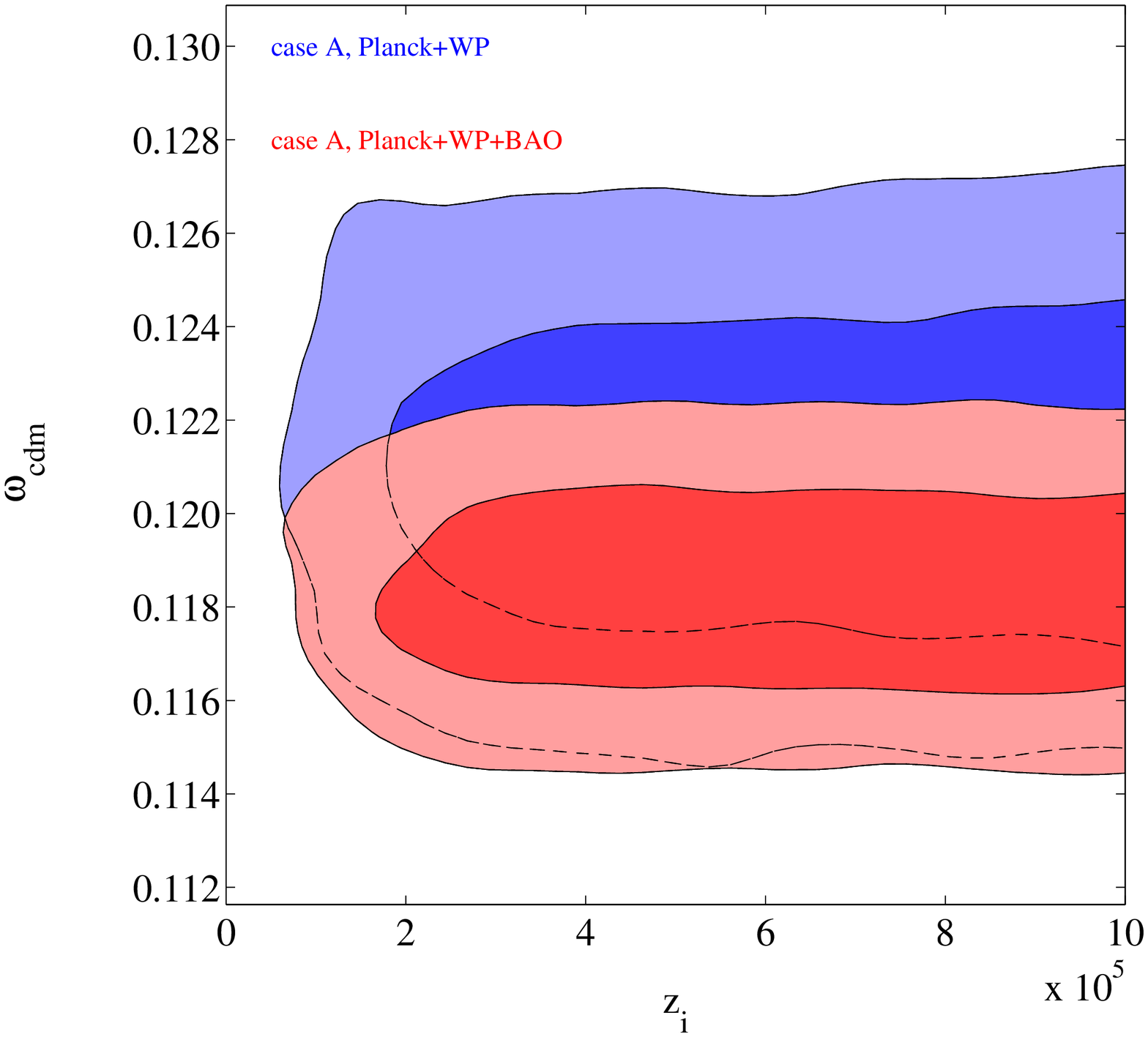}&\includegraphics[scale=0.3]{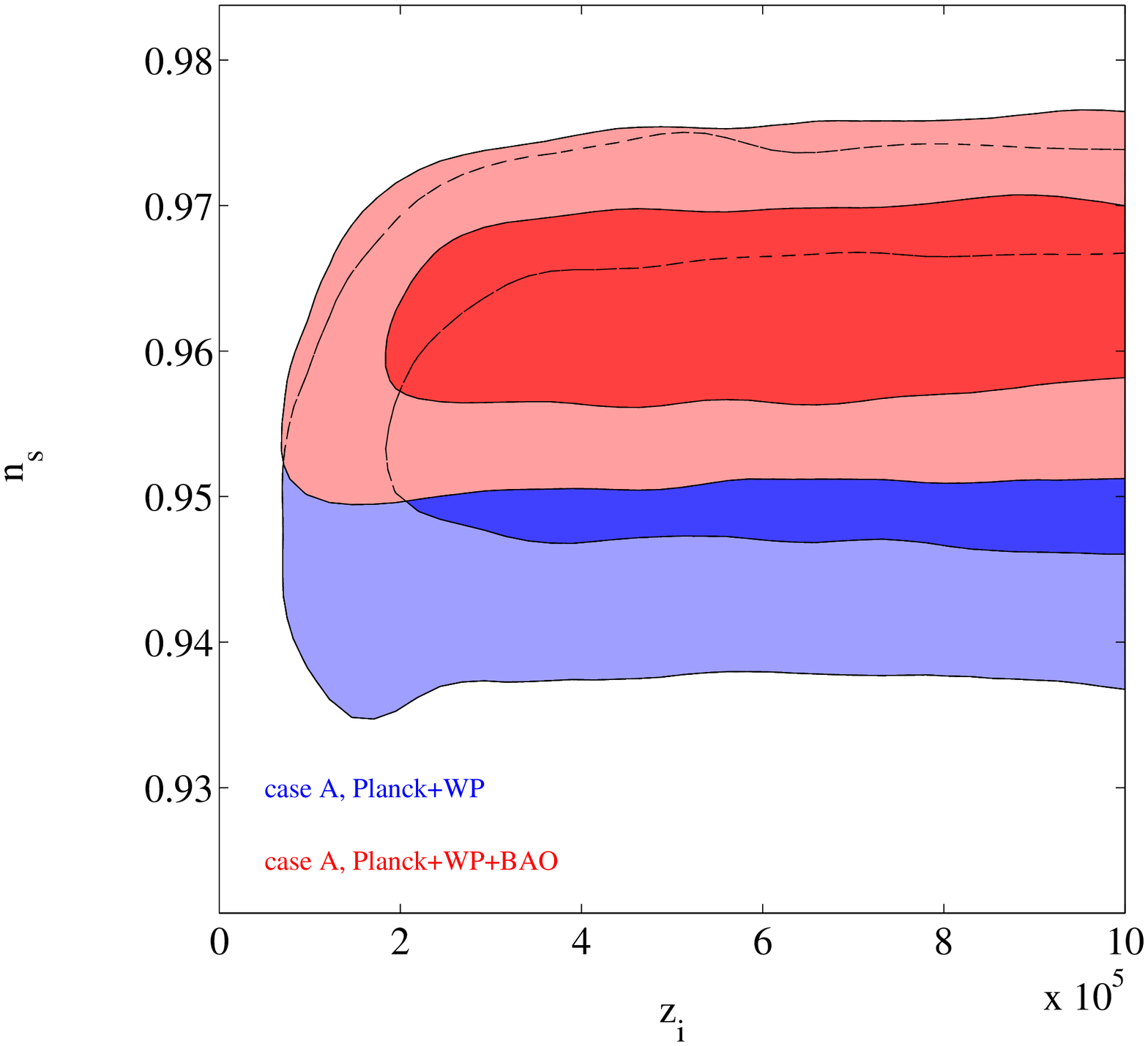}\\
\includegraphics[scale=0.3]{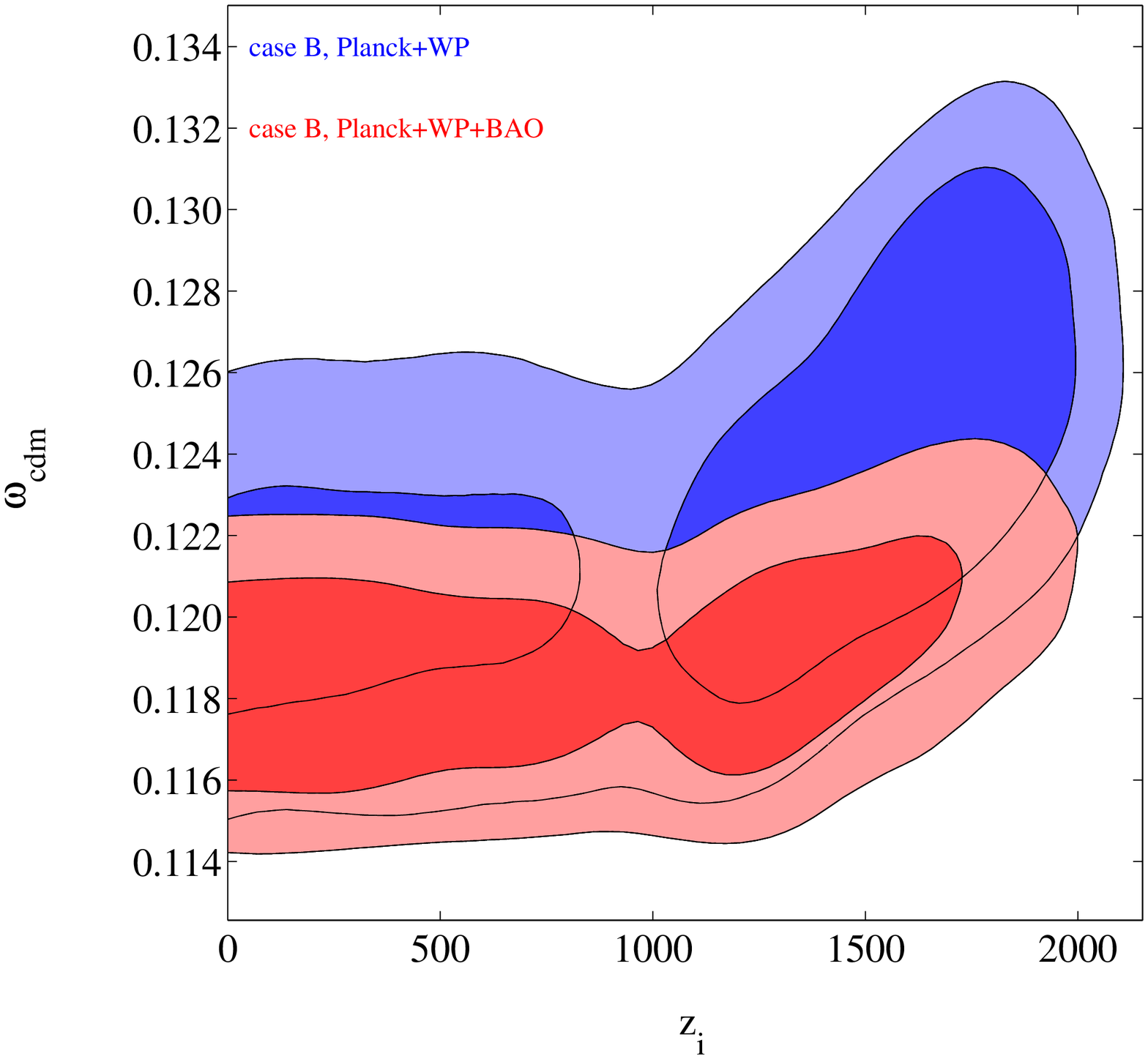}&\includegraphics[scale=0.3]{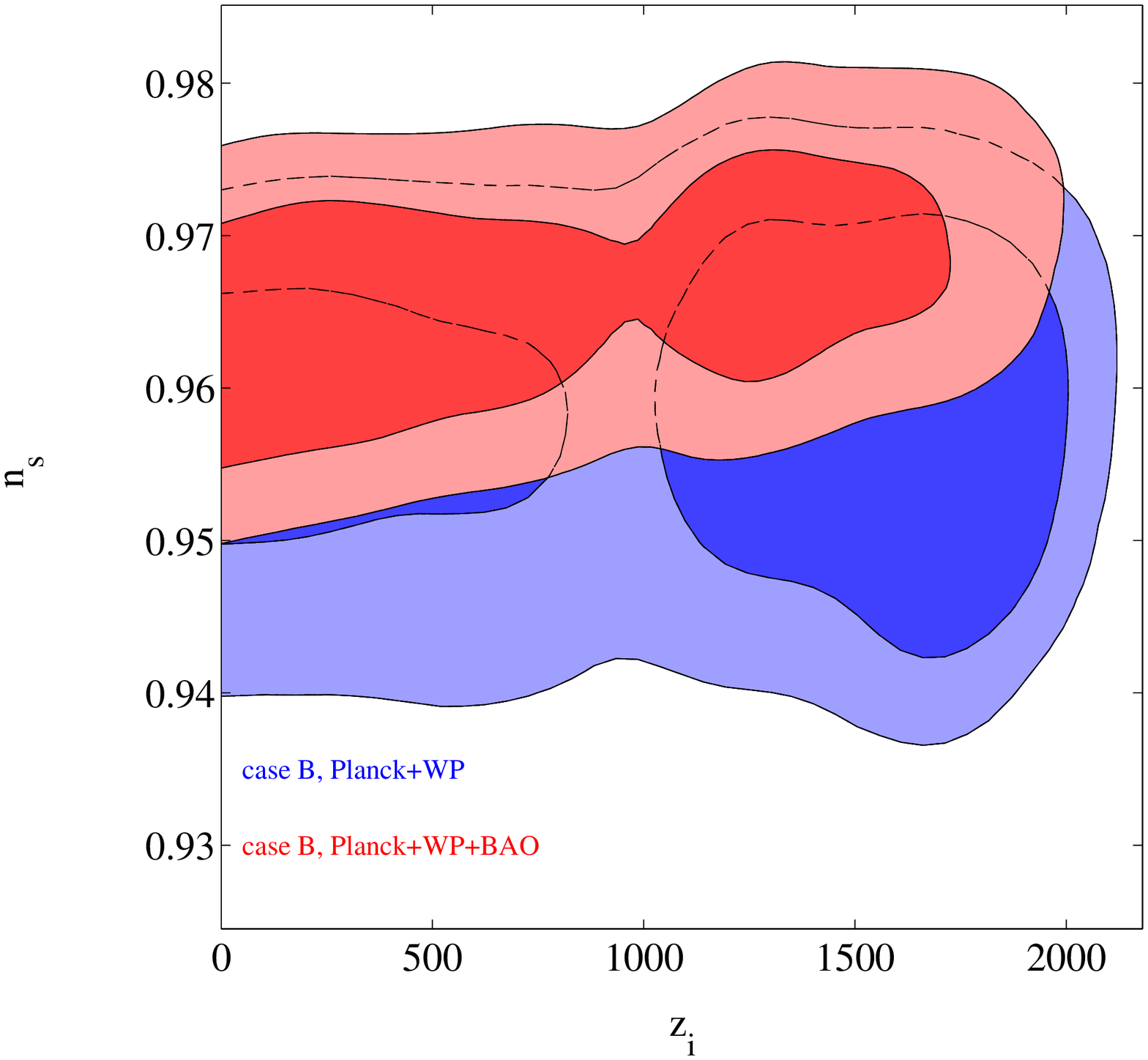}\\
\end{tabular}
\caption{2D marginal 68\% and 95\% contours for $z_i$ and $\omega_{\rm cdm}$ (left panels), $z_i$ and $n_{s}$ (right panels), for case A (top panels) and case B (bottom panels).}
\label{fig:correlations}
\end{figure}

\begin{figure}[h]
\centering
\includegraphics[scale=0.45]{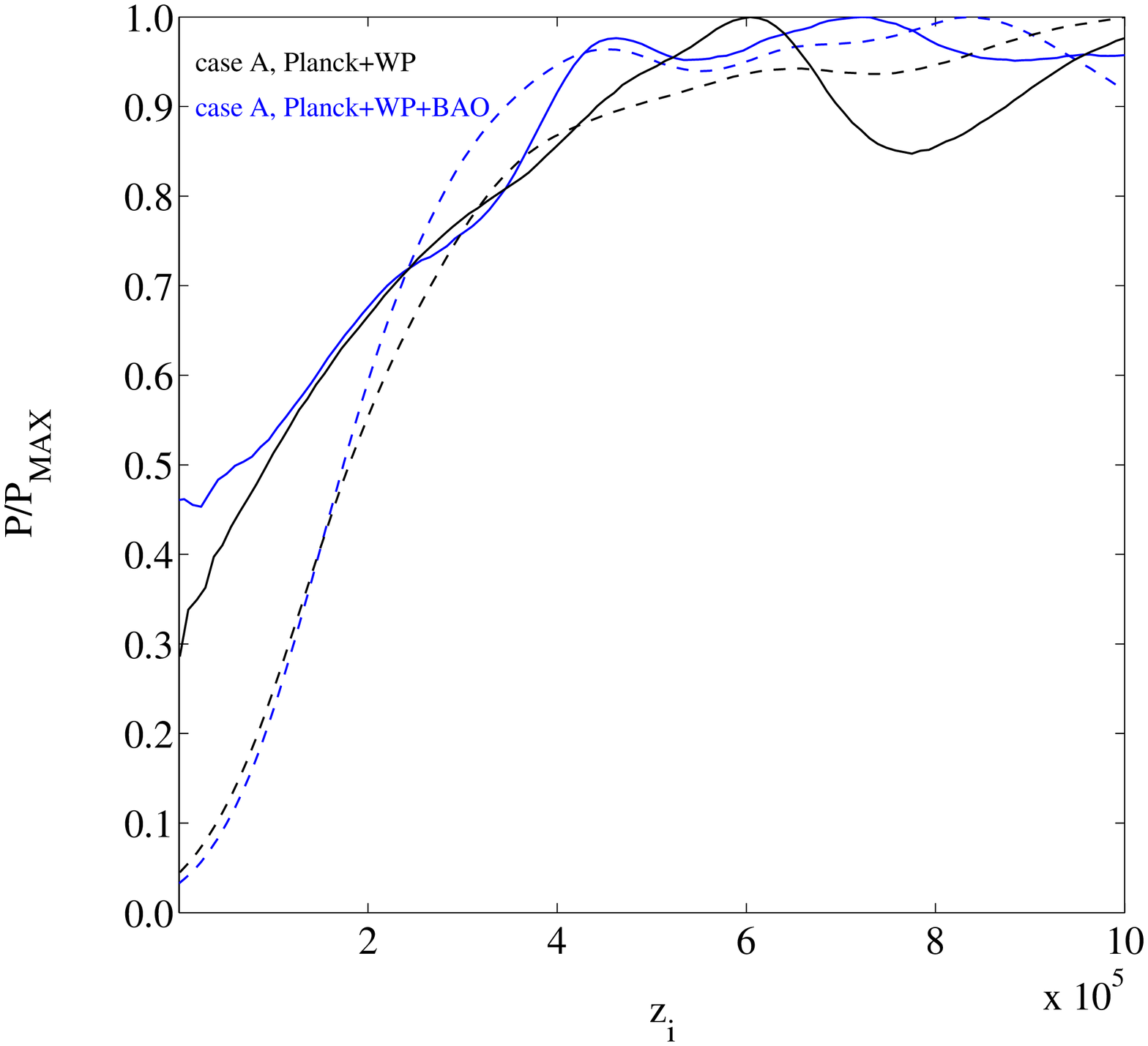}\\
\includegraphics[scale=0.45]{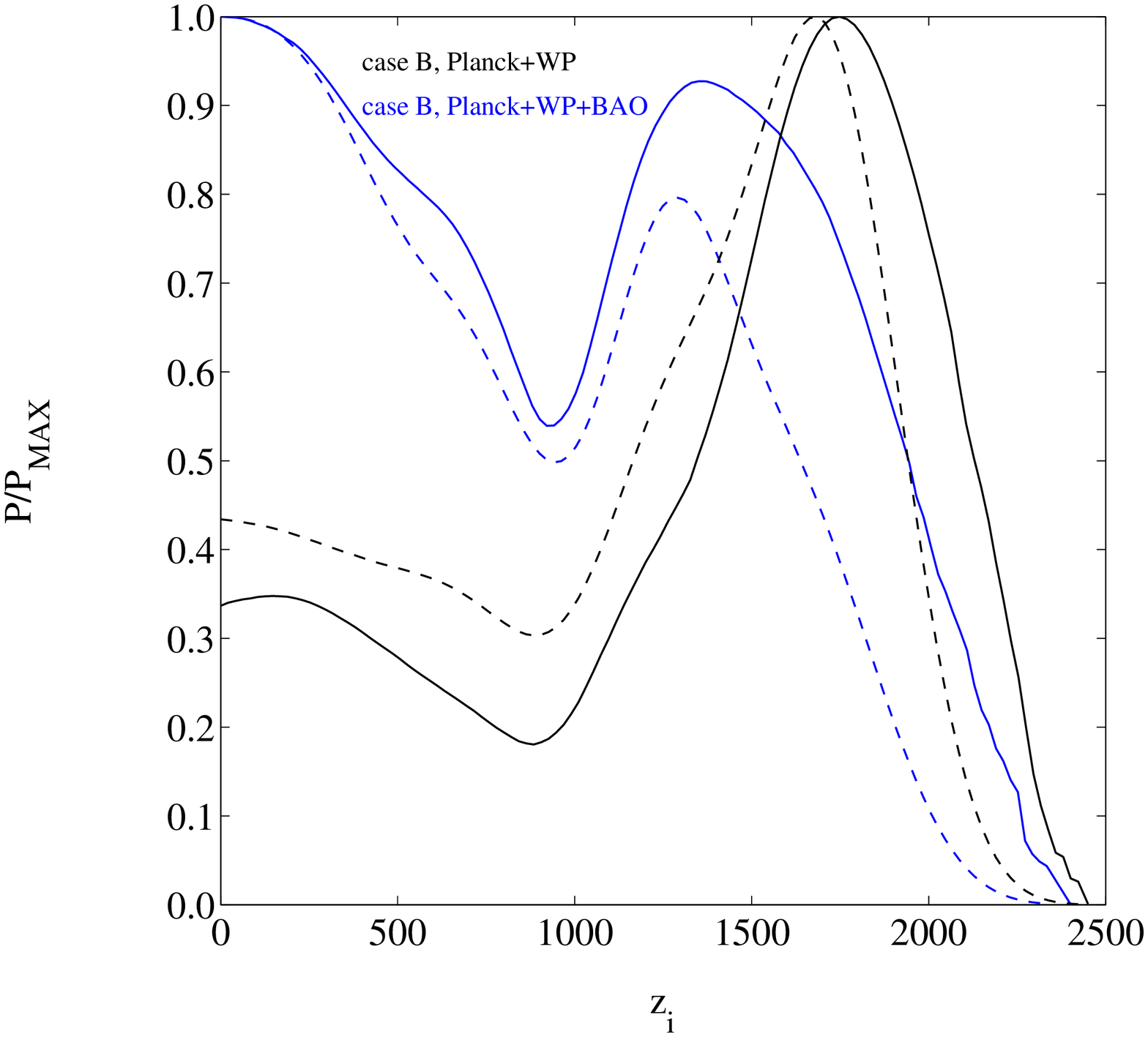}
\caption{One dimensional marginalized posteriors (solid line) and mean likelihoods (dotted lines) derived in the Fermi like model (upper panel) and in the pseudoscalar model (lower panel). The black lines refer to the analyses including only Planck+WP, while the blue lines include also BAO.}
\label{fig:zi}
\end{figure}

\begin{figure}[h]
\centering
\includegraphics[scale=0.45]{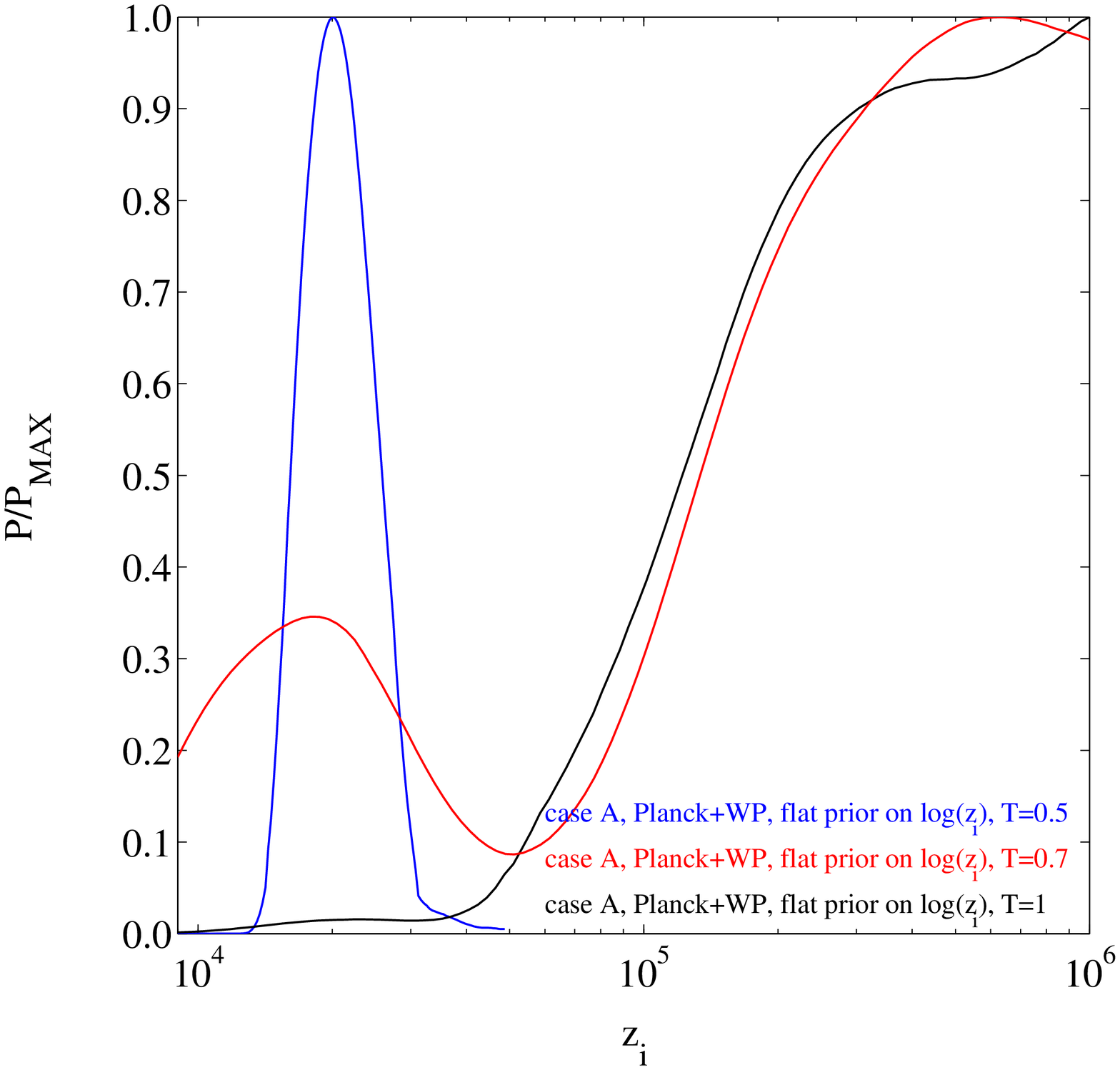}
\caption{One dimensional marginalized posteriors derived in the Fermi like model assuming a flat prior on $\log(z_i)$ and varying the temperature of the MCMC.}
\label{fig:logzi}
\end{figure}

\section{Cosmological model}
\label{sec:model}

Our analysis is carried out in the framework of the $\Lambda$CDM model, but extended to include massive neutrinos, and with additional parameters describing their interactions.
The set of cosmological parameters is therefore given by
\begin{equation}\label{eq:model}
{\bm \theta} = \{\omega_{\rm cdm},\omega_{\rm b},\theta_{\rm s},\tau,
\ln(10^{10}A_{s}),n_{s}, \Sigma m_\nu, {\cal I}\}.
\end{equation}
Here, $\omega_{\rm cdm} \equiv \Omega_{\rm cdm} h^2$ and $\omega_{\rm b}
\equiv \Omega_{\rm b} h^2$ are the present-day physical CDM and baryon densities
respectively, $\theta_{\rm s}$ the angular size of the sound horizon, $\tau$  the optical
depth to reionization, and $\ln(10^{10}A_s)$ and $n_s$ denote respectively the amplitude and spectral index
of the initial scalar fluctuations.  
The neutrino sector is characterized by three neutrinos sharing the same mass parameterized through the mass sum $\Sigma m_\nu$, and a vector of new interaction parameters, ${\cal I}$. In the present case where
we approximate the interaction with a redshift of transition between non-interacting and extremely strongly interacting regimes we simply have ${\cal I} = z_i$. In Case A,  Fermi-like 4-point interaction the additional parameter $z_i$ is spanning the high redshift range up to $z_i\sim10^7$ with both a flat and a logarithmic prior on it. In case B, interaction mediated by a pseudo-scalar, we assume a flat prior on $0<z_i<10^4$. In order to further discuss this approximation we will perform an analysis where we account for the collisional Boltzmann equations and we use the coupling as additional parameter (${\cal I} = \log\left(G_X\,{\rm MeV}^2\right)$ in case A and ${\cal I} = \log(g)$ in case B) instead of $z_i$ and we assume a flat prior on the logarithm of the coupling.

\begin{table}[t]
\caption{Priors for the cosmological fit parameters considered which are not related to neutrino interactions. All priors are uniform (top hat) in the given intervals.}
\label{tab:priors}
\begin{center}
\begin{tabular}{lc}
\hline
 Parameter & Prior\\
\hline
$\omega_{\rm b}$ & $0.005 \to 0.1$\\
$\omega_{\rm cdm}$ & $0.001 \to 0.99$\\
$\theta_{\rm s}$ & $0.5 \to 10$\\
$\tau$ & $0.01 \to 0.8$\\
$n_s$ & $0.9 \to 1.1$\\
$\ln{(10^{10} A_s)}$ & $2.7 \to 4$\\
$\Sigma m_\nu$ [eV] &  $0 \to 3$\\
\hline
\end{tabular}
\end{center}
\end{table}

\section{Data and analysis}
\label{sec:data}

We consider two types of measurements: temperature and polarization power spectra of the CMB anisotropies and the Baryonic Acoustic Oscillations (BAO).  These are discussed in more detail below.  To these data sets we apply a Bayesian statistical inference analysis using the publicly available Markov Chain Monte Carlo
parameter estimation package {\sc CosmoMC}~\cite{Lewis:2002ah} coupled to the CAMB~\cite{Lewis:1999bs} Boltzmann solver modified to accommodate the two versions of non-standard neutrino interactions.  The likelihood routines and the associated window functions are supplied by the experimental collaborations.

\subsection{CMB anisotropies}

Our primary data set is the recent measurement of the CMB temperature (TT) power spectrum by the Planck mission~\cite{Planck:2013kta}, which we implement into our
likelihood analysis following the procedure reported in~\cite{Ade:2013zuv}.
This data is supplemented by measurements of the CMB polarization from the WMAP nine-year data release~\cite{Bennett:2012fp}, in the form of
an autocorrelation (EE) power spectrum at $2<\ell<32$ and a cross-correlation (TE) with the Planck temperature measurements in the same multipole range.  We denote this supplement ``WP''.

\subsection{Auxiliary data}

In addition we include the information on the Baryonic Acoustic Oscillations (BAO) extracted from three different galaxy redshift surveys: the Data Release 7 (DR7) of the Sloan Digital Sky Survey (SDSS) at redshift $z = 0.35$ \cite{Padmanabhan:2012hf}, Data Release 9 (DR9) of the Baryon Acoustic Spectroscopic Survey (BOSS) measurement at $z = 0.57$ \cite{Anderson:2012sa} and the 6dF Galaxy Survey at $z = 0.1$ \cite{Beutler:2011hx}.

\section{Results}
\label{sec:results}

\subsection{4-point interactions}

In Figs.~\ref{fig:clscaseA} and \ref{fig:clsratiocaseA} we show power spectra as well as their ratio with respect to the $\Lambda$MDM model for various values of $z_i$. Exactly as expected there is almost no change relative to $\Lambda$MDM for small values of $\ell$. However, for larger $\ell$ the difference becomes very pronounced. The transition between the two regimes corresponds roughly to the sound horizon size at $z_i$. Given that the comoving sound horizon scale is $3^{-1/2} aH \propto a^{-1}$ in the radiation dominated epoch we can very approximately assume that the $\ell$-value where interactions become important is around $\ell \sim 4\times10^{-2}z_i$. We indeed see exactly this effect in Fig.~\ref{fig:clsratiocaseA}. The $\ell$ corresponding to the middle of the transition region approximately corresponds to the sound horizon at $z_i$.
Our MCMC runs yield a lower limit on $z_i$ of  $z_i > 1.8\times10^5$ at 95\% c.l. when only Planck+WP data is used, a result which changes only marginally to 
$z_i > 1.9\times10^5$ at 95\% c.l. when BAO data are included. 
Concerning the neutrino mass sum the $2~\sigma$ upper bounds are $\Sigma m_\nu<0.90$~eV using only Planck+WP and $\Sigma m_\nu<0.24$~eV including also BAO. As we can see in Fig.~\ref{fig:correlations} (upper panels), there is no strong correlation between $z_i$ and the ``vanilla'' parameters; this means that the theoretical power spectrum shown in Fig.~\ref{fig:clscaseA} cannot be reproduced by the variation of other cosmological parameters mimicking massive neutrinos strongly interacting up to redshift $z_i$. The posteriors in Fig.~\ref{fig:zi} (upper panel) show a rapid increase with increasing $z_i$, completely as expected. The Planck+WP data show a peak around $z_i\sim6\times10^5$. However, this peak has no statistical significance and it is a spurious effect due to the paucity of the sampling in the high redshift region. The bounds on the redshift of the decoupling can be translated into a bound on the coupling constant $G_X$ involved in the Fermi like-4 point interaction; using Eq.~\ref{eq:gtozcaseA} we obtain $G_X \leq \left(0.06 \, {\rm GeV}\right)^{-2}$, corresponding to $G_X \leq 2.5 \times 10^7 G_F$. 
If the approximation of the instantaneous decoupling is discarded and the collisional Boltzmann equations are used with the logarithm of the coupling in ${\rm MeV}^2$ as free parameter, then the coupling turns out to be $\log_{10}(G_X{\rm MeV}^2) \leq -3.9$ at 95\% c.l..
This bound corresponds $G_X \leq \left(0.08 \, {\rm GeV}\right)^{-2}$ at 95\% c.l., which is slightly tighter than the result found under the assumption of the instantaneous approximation.
Furthermore this limit seems consistent with the standard mode of Ref.~\cite{Cyr-Racine:2013jua} where a bound of $\log_{10}(G_X{\rm MeV}^2) \leq -3.5$ is quoted at 95\% confidence.  
With uniform priors on $z_i$ we see no evidence of the peak in the posterior likelihood around $z_i \sim 10^4$ found in Ref.~\cite{Cyr-Racine:2013jua}.
However, when we use uniform priors on $\log (z_i)$ and we lower the temperature of the MCMC, we have verified that we reproduce this peak (Fig.~\ref{fig:logzi}). Given the instability of this result with respect to the prior, we conclude that the current data cannot adequately discriminate between the two regions ($z_i \sim 10^4$ and $z_i > 1.8 \times 10^5$) of the parameter space. Furthermore Fig.~\ref{fig:logzi} shows how the constraints depend not only on the prior but also on the temperature of the monte carlo and on the starting point: basically if the temperature is too low and the starting point is around $z_i=10^4$, then the sampling gets stuck in the local maximum of the likelihood around $z_i \sim 10^4$, on the contrary, starting from the same point, if the temperature is too high, then the sampling is not sensitive to the peak.

Finally it should be emphasized that neutrinos strongly interacting at all times ($z_i \to 0$) are highly disfavored by CMB data ($\Delta\chi^2\sim10^4$), i.e.\ the case of no neutrino anisotropic stress is very highly 
disfavored by current data (see \cite{Bashinsky:2003tk,Hannestad:2004qu,Trotta:2004ty,Bell:2005dr,Sawyer:2006ju,Friedland:2007vv,Smith:2011es} for earlier treatments of the question of neutrino anisotropic stress).

\subsection{Pseudoscalar interactions}

In Figs.~\ref{fig:clscaseB} and \ref{fig:clsratiocaseB} we show, respectively, the power spectra and the ratio respect to the $\Lambda$MDM model for various values of $z_i$. The expectation here is as follows: No modification on scales larger than the sound horizon at recombination. An increase in power for $\ell$ smaller than the value corresponding to the sound horizon at $z_i$, i.e.\ for $\ell \leq 5 z_i^{1/2}$.
The recoupling redshift is lower than 1887 at 95\%c.l. for Planck+WP ($1737$ for Planck+WP+BAO).
The constraints on the sum of neutrino masses are consistent with those obtained in case A): $\Sigma m_\nu<0.95$~eV at 95\%c.l. using Planck+WP and $\Sigma m_\nu<0.25$~eV at 95\%c.l. using Planck+WP+BAO. We can notice that the bounds we get on $\Sigma m_\nu$ justify the single fluid assumption (see section \ref{sec:pseudoscalarinteractions}). In Fig.~\ref{fig:correlations} (lower panels) we can appreciate a correlation between $z_i$ and $\omega_{cdm}$. The Planck+WP posterior (black solid line) in Fig.~\ref{fig:zi} (lower panel) shows a bimodal distribution with a peak around the best-fit value $z_i=1600$.  We have checked the presence of this likelihood maximum by sampling the probability distribution in the parameter space with a higher temperature parameter ($T=2$ rather than $T=1$) \cite{Hamann:2011hu}. We also analyze separately this nearly Gaussian peak of the posterior distribution and find that the recoupling redshift is lower than 1965 at 95\%c.l.

\paragraph{Translating into a bound on $g$}

Using the same argument as in Ref.~\cite{Basboll:2008fx} we can translate $z_i$ into a bound on the dimensionless coupling. Since $z_i$ is constrained to be relatively small we can use the expression given in Eq.~(\ref{eq:gtozcaseB}) for the matter dominated epoch
to derive a bound of approximately
\begin{equation}
g_{ii}<1.2\times10^{-7}
\end{equation}
at 95\% confidence, applicable to both the Planck data only and Planck+auxiliary data cases.
In the analysis where the interactions are described by the collisional Boltzmann equations and the free parameter is the coupling constant with a flat prior on the logarithm of $g$, the 95\% upper bound is
\begin{equation}
g_{ii}<1.5\times10^{-7}\,,
\end{equation}
consistent with the result of the approximated case. Furthermore we notice that the latter limit is slightly higher than in the approximated case, as we expected from considering the bias introduced by our approximation (see Fig.~\ref{fig:clstaupseudo}).

Because of the steep dependence of this bound on $z_i$, the limit here is formally only marginally different from the limits presented 
in \cite{Basboll:2008fx}.

In the same way the bound on the off-diagonal elements is given approximately by
\begin{eqnarray}
g&<&0.6\times10^{-11} \left(\frac{1+z_{\rm dec}}{1088+1}\right)^{9/4}\left(\frac{50\textrm{ meV}}{m}\right)^2\, ,
\end{eqnarray}
so that the upper bound is approximately 
\begin{eqnarray}
g&<&2.3\times10^{-11}\left(\frac{50\textrm{ meV}}{m}\right)^2.
\end{eqnarray}
The latter limit can also be translated into a bound on the lifetime in the neutrino rest frame of
\begin{eqnarray}
\tau&>&1.20\times10^{9}\textrm{ s }\left(\frac{m}{50\textrm{ meV}}\right)^3\,.
\end{eqnarray}
If we discard the assumption about the matter domination by including the full redshift dependence of $H(z)$, the lifetime is slightly higher
\begin{eqnarray}
\tau&>&1.24\times10^{9}\textrm{ s }\left(\frac{m}{50\textrm{ meV}}\right)^3\,.
\end{eqnarray}

The limits on the neutrino decay lifetime are weaker than those reported in Ref.~\cite{Basboll:2008fx}
because the bounds on $z_i$ are broader and the dependence of the Yukawa off-diagonal elements on $z_i$ is not as steep as in the case of the diagonal elements.

Finally, we again find that the case of neutrinos which are strongly interacting at all times ($z_i \to \infty$ in this case) is very highly
disfavored with $\Delta\chi^2\sim10^4$.

\section{Discussion}
\label{sec:discussion}
We have investigated the impact on the Cosmic Microwave Background of two different versions of non-standard neutrino interactions: a Fermi-like 4 point interaction and a pseudoscalar interaction mediated by the Nambu-Goldston boson of a broken $U(1)$ symmetry.

The former case leads to strong neutrino self-interactions at high temperature which subsequently decouple as the Universe expands and the temperature decreases.
The effect on the CMB power spectrum is an increase in power on scales smaller than the scale of the sound horizon at the transition to the free-streaming regime. Current cosmological data (CMB data from Planck and WMAP as well as BAO data) allow for the presence of a non-standard 4-point interaction provided that neutrino self-interactions decouple at a redshift of $z_i \sim 10^5$ \footnote{Besides this high redshift mode, the posterior shows a peak around $z_i\sim10^4$ when a uniform prior on $\log(z_i)$ is assumed; however additional data are needed in order to assess the significance of this peak.}, corresponding to an effective coupling strength of $G_X \sim (0.06 \, {\rm GeV})^{-2}$.

The latter case, the pseudoscalar interaction, has the opposite behavior in the sense that neutrino self-interactions become important at low redshift such that neutrinos become 
strongly interacting at some redshift $z_i$, through a combination of binary processes, decays, and inverse decays.
In this case the effect on the CMB power spectrum is located at multipoles lower than the multipole corresponding to the sound horizon when these interactions become important. 
We note that although very strong self-interacting which equilibrate neutrinos prior to $z \sim 1900$ are strongly disfavored there seems to be a mild preference for an equilibration redshift around $z_i \sim 1500$. Although this in principle could be pointing to the presence of neutrino self-interactions the significance is quite low and here we simply quote the upper bound on $z_i$.

%In accordance with previous studies we find that strongly interacting neutrinos are highly disfavored by %present data ($\Delta \chi^2 \sim 10^4$), showing that neutrino free-streaming is an absolutely necessary %ingredient of standard cosmology.
Although we have found strongly interacting neutrinos to be compatible with cosmology provided they decouple early enough or recouple late enough, in accordance with previous studies, we conclude that neutrinos that are strongly interacting throughout all cosmic history are highly disfavored by present data ($\Delta \chi^2 \sim 10^4$). Neutrino free-streaming is a necessary ingredient of standard cosmology.

Finally we note that the cosmological neutrino mass bound does not depend strongly on neutrino self-interactions {\it provided} that the mass is low enough that neutrinos are still  relativistic around the epoch of recombination.

\section*{Acknowledgements}

The authors acknowledge the European ITN project Invisibles (FP7-PEOPLE-2011-ITN, PITN-GA-2011-289442- INVISIBLES).

We would like to thank the anonymous referee for comments and suggestions which helped to improve the manuscript.

\bibliographystyle{utcaps}

%\bibliography{refs}

\providecommand{\href}[2]{#2}\begingroup\raggedright\endgroup

\end{document}